\documentclass[aps,pra,reprint,superscriptaddress]{revtex4-2}

\usepackage{graphicx}
\usepackage{amsmath}
\usepackage{amssymb}
\usepackage{xcolor}

\definecolor{darkblue}{rgb}{0,0,.65}
\definecolor{darkgreen}{rgb}{0.2,0.6,0.2}
\usepackage[%
pdfstartview=FitH,%
breaklinks=true,%
bookmarks=true,%
colorlinks=true,%
anchorcolor=black,%
citecolor=darkgreen,
filecolor=black,%
menucolor=black,%
urlcolor=darkblue,%
linkcolor=blue,%
]{hyperref}

\begin{document}

\title{Dynamics of quasiholes and quasiparticles at the edges of small lattices}
	
\author{Xikun Li}
\affiliation{School of Physics and Optoelectronic Engineering, Anhui University, Hefei, Anhui 230601, China}
\affiliation{Max-Planck-Institut f\"{u}r Physik komplexer Systeme, D-01187 Dresden, Germany}
\author{B{\l}a\.{z}ej Jaworowski}
\affiliation{Department of Physics and Astronomy, Aarhus University, Ny Munkegade 120, DK-8000 Aarhus C}
\author{Masudul Haque}
\affiliation{Institut f{\"u}r Theoretische Physik, Technische Universit{\"a}t Dresden, D-01062 Dresden, Germany}
\affiliation{Max-Planck-Institut f\"{u}r Physik komplexer Systeme, D-01187 Dresden, Germany}
\affiliation{Department of Theoretical Physics, Maynooth University, County Kildare, Ireland}
\author{Anne E. B. Nielsen}
\affiliation{Department of Physics and Astronomy, Aarhus University, Ny Munkegade 120, DK-8000 Aarhus C}
\affiliation{Max-Planck-Institut f\"{u}r Physik komplexer Systeme, D-01187 Dresden, Germany}

\begin{abstract}
We study quench dynamics of bosonic fractional quantum Hall systems in small lattices with cylindrical boundary conditions and low particle density.  The states studied have quasiholes or quasiparticles relative to the bosonic Laughlin state at half filling. Pinning potentials are placed at edge sites (or sites close to the edges) to trap quasiholes and qausiparticles. The potentials are then turned off, and because the edges of fractional quantum Hall systems host chiral edge modes, we expect chiral dynamics of the quasiholes and quasiparticles. We numerically show that chiral motion of the density distribution is observed and robust for the case with positive potentials (quasiholes), but that there is no noticeable chiral motion for negative potentials (quasiparticles). The comparison of the numerical ground states with model lattice Laughlin wavefunctions suggests that both positive and negative potentials do create and pin anyons that are not necessarily well-separated on small lattices. Initializing the dynamics with the model state also shows the lack of chiral dynamics of quasiparticles.  Our results suggest that, in small lattices with low particle density, quasiparticles are strongly adversely affected in dynamical processes, whereas quasiholes are dynamically robust. 
\end{abstract}

\maketitle

\section{Introduction}
A fascinating aspect of topologically ordered phases of matter is that they support anyonic quasiparticles with fractional exchange statistics \cite{Wilczek82}. Anyons appear, e.g., as charged excitations of fractional quantum Hall (FQH) systems in two-dimensional electron gases subject to strong magnetic fields \cite{Laughlin1983}. The robustness of exchange statistics against local noise makes anyons an interesting platform for topological quantum computation, which has motivated much work towards realizing anyons in different physical systems \cite{Kitaev2003,Nayak08}. Various models hosting FQH physics and anyons have been proposed for different physical platforms~\cite{Du2009, Bolotin2009, Regnault2011,Wang2014,Lohse2018,Nandkishore2019}. Both the fractional charge~\cite{Goldman1995,Picciotto1997,Saminadayar1997,Martin2004} and the fractional statistics \cite{Nakamura2019} of anyons have been observed experimentally. 
	
Ultracold atoms in optical lattices provide a versatile setup to study collective quantum physics \cite{Bloch2005,Bloch2008,Aidelsburger2013,Ott2016}. Several lattice models displaying FQH physics and schemes for implementations in ultracold atoms have been proposed \cite{Paredes2003,Sorensen2005,Hafezi2007,Neupert2011,JuliaDiaz2011,Hormozi2012,Cooper2013,He2017}, and recently a FQH system with two particles on 16 sites was realized experimentally \cite{leonard2022}. Moreover, it has been claimed that anyons can be created and trapped with pinning potentials at specific positions \cite{Storni2011,Johri2014,Raciunas2018,Wang22}, and that the braiding of anyons can be realized by adiabatically moving the potentials \cite{Kapit2010,macaluso2020}.
	
It is a crucial question whether the characteristic features of FQH systems in two-dimensional electron gases still hold for small lattice systems with low particle densities, or what features should be chosen to characterize FQH states in such systems~\cite{Raciunas2018,Wang22}. Small lattice systems are ideal testing grounds from both experimental and computational viewpoints: Experimentally, the manipulation of single sites and single atoms~\cite{Ott2016,Parsons2015} provides tools to observe the FQH effect in small lattices (if we know what to observe), and computationally, the exponentially growing size of the Hilbert space puts a limit on the system sizes that can be studied. Proposals for probing the fractional charge and fractional statistics in small optical lattices have been made \cite{Kapit2010,Raciunas2018,macaluso2020,Srivatsa2021}. It is, however, an open question whether known results for solid-state systems with macroscopic numbers of electrons are still true for small (or even medium) size lattice systems with dozens of sites and few particles. Finite-size effects can influence the results in various ways, for example by making it impossible to separate the anyons sufficiently, or by altering the energy spectrum. Another issue is that, due to low site occupancies, the process of pinning or localizing quasiparticles could excite an uncontrolled amount of excitations.    

One of the characteristic features of FQH states is the presence of chiral modes at the edges. This result was initially derived for large continuum systems (long-wavelength limit) \cite{wen1990chiral}, but numerical calculations have shown chiral motion of quasiholes in small continuum systems \cite{grass2012quasihole,li2022anyonic} and chiral motion of ``full'' particles in (relatively large) lattice systems \cite{dong2018charge}. In addition, even very small lattice systems retain some degree of similarity to the continuum, which is seen in the counting of the edge states \cite{binanti2023edge}. It is therefore relevant to ask whether the chiral motion of charges such as quasiparticles and quasiholes can be observed in small lattice systems, comparable to the ones used in experiments \cite{leonard2022}.

Here, we investigate whether quasiholes (quasiparticles) can be pinned by positive (negative) potentials in small lattices and whether chiral dynamics along the edge is present in the quench dynamics (i.e.\ after turning off the pinning potential). We consider hardcore bosonic FQH systems with filling factor $\nu=1/2$ and place pinning potentials at or near the edges of the lattice. The hardcore constraint is the $U/J \rightarrow \infty$ limit, where $U$ is the on-site interaction, and $J$ is the hopping strength. In practice, when $U/J$ is large enough, the hardcore constraint is a good approximation (for negative pinning potentials $V_0$, we also need $|V_0|\ll U$). This results in the creation of density depletions (increases), which we interpret as quasiholes (quasiparticles) due to the accumulated excess density being close to $\pm0.5$ and due to high overlap with model wavefunctions. For the lattice sizes considered in this work, i.e.\ dozens of sites, we observe chiral motion of quasiholes. We also find that the chiral motion is robust in the sense that it is observed for a range of lattice sizes and for various strengths and locations of the potentials. On the contrary, we do not observe chiral dynamics for the quasiparticles.

We find that the ground states with potentials (both positive and negative) have high overlaps with model lattice Laughlin states with anyons, suggesting that in both cases we do create anyons.  This suggests that the  absence of chiral motion for negative potentials stems from complexities arising in the dynamics of quasiparticles, presumably involving a large number of highly excited states.  Thus, we show that in these setups (small lattices and low particle densities) there are substantial differences between the dynamics of quasiparticles and the dynamics of quasiholes. 
	
The paper is structured as follows. In Sec.~\ref{sec:model}, we introduce the model. In Sec.~\ref{sec:hole}, we consider the case with positive potentials and demonstrate chiral motion of the density distribution in the quench dynamics. In Sec.~\ref{sec:electron}, we study the case with negative pinning potentials and find no chiral motion of the density distribution. In Sec.~\ref{sec:modelstate} we compare the ground state of the Hamiltonian with potentials with model lattice Laughlin states with anyons, showing that the overlap is high.  In Sec.~\ref{sec:conclusions} we expand on the proposed explanation of our main observations and conclude the paper by pointing out implications of our results. Appendix \ref{app:inversion} explains the model lattice wavefunctions used in Sec.~\ref{sec:modelstate}, and Appendix \ref{app:LatticeVsContinuum} shows how to relate them to continuum Laughlin wavefunctions.

\section{Model}\label{sec:model}
	
We consider an interacting Hofstadter model with hardcore bosons on a two-dimensional square lattice. The sites are at the positions $\vec{r}_i=a(x_i,y_i)$, where $a$ is the lattice constant, $x_i\in\{1,2,\ldots,N_x\}$, and $y_i\in\{1,2,\ldots,N_y\}$. The Hamiltonian takes the form
\begin{align}\label{eq:BH_1}
		H_0 = -  J\sum_{\langle k,j \rangle}  c^{\dagger}_{k} c_{j} e^{ \frac{2 \pi i}{\phi_0} \int^{\vec{r}_k}_{\vec{r}_j} \vec{A} \cdot \mathrm{d}\vec{r}},
\end{align}
where the sum is over all $k$ and $j$ for which the sites at $\vec{r}_k$ and $\vec{r}_j$ are nearest neighbors. Note that each pair of neighbors appears twice in the sum. The operator $c_{k}$ annihilates a boson at site $\vec{r}_k$, $\phi_0=h/e$ is the flux quantum, $h$ is Planck's constant, $e$ is the elementary charge, and $\vec{A}$ is the vector potential corresponding to a uniform magnetic field $B$ perpendicular to the plane of the lattice. The strength of the nearest neighbor hopping is set to $J=1$ for simplicity.  The interaction between bosons is included implicitly by enforcing the hardcore condition, i.e.\ allowing at most one boson per site.  
	
We consider a square lattice with open boundary conditions in the $x$ direction and periodic boundary conditions in the $y$ direction.  This cylinder topology provides two parallel edges without corners, and hence is particularly suitable for studying chiral dynamics along the edges. We use the Landau gauge
\begin{equation}
	\vec{A}=B(0,x-x_0,0)
	\label{eq:landaugauge}
\end{equation}
where $x_0=(N_x+1)/2$, which ensures that the vector potential vanishes at the center of the system. Note that $a^2B=\alpha\phi_0$, where $\alpha$ is the number of flux quanta per plaquette. 
	
We consider half filling $\nu=1/2$ in this paper, where the filling factor $\nu$ is the number of particles divided by the number of flux quanta penetrating the lattice. (Note that $\nu=1/2$ is the filling fraction of the lowest Hofstadter band, and not the lattice filling fraction or average particle density, which is considerably smaller in our case.)  If there are quasiholes (quasiparticles) present, they each count for $\nu$ particles ($-\nu$ particles) when computing $\nu$. The number of flux quanta is $\alpha$ times the number of plaquettes $N_{\mathrm{plaq}}$ in the lattice. Altogether, we hence have
\begin{align}\label{eq:filling_plaq}
	\nu=\frac{M+\nu \sum_k p_k}{\alpha N_{\mathrm{plaq}}},
\end{align}
where $M$ is the number of particles and $p_k=+1$ ($p_k=-1$) if the $k$th anyon is a quasihole (quaisparticle). At the edge appearing due to the open boundary conditions in the $x$ direction, there are ambiguities in how to count the amount of flux through the lattice, and this ambiguity is significant for small lattices. It was suggested in \cite{Raciunas2018} that the choice $N_{\mathrm{plaq}} = (N_x-1) N_y$ is best for creating and stabilizing a FQH droplet in small lattice systems, and we hence also use this way of counting the flux here. We compute $\alpha$ from \eqref{eq:filling_plaq} with $\nu=1/2$, and this gives us the vector potential $\vec{A}$ appearing in the Hamiltonian.
	
Since we are considering the $\nu=1/2$ Laughlin state, a quasihole corresponds to half a particle missing in a local region, and a quasiparticle corresponds to half a particle extra in a local region. We would like the anyons to be pinned at particular positions to begin with and local pinning potentials are a standard approach to do that. Adding a potential term $V_jn_j$ to the Hamiltonian, where $n_j=c_j^\dag c_j$ is the particle number operator, results in an energy penalty for a particle to sit at the $j$th site if $V_j>0$ and the opposite for $V_j<0$. A natural way to try to trap two quasiholes is hence to reduce the number of particles $M$ by one and add positive potentials on two sites. Similarly, one could try to trap two quasiparticles by increasing the number of particles $M$ by one and putting negative potentials on two sites. Note that $M+\nu\sum_kp_k$ is unchanged in this process, and hence $\alpha$ is also unchanged. The Hamiltonian with potentials is
\begin{equation}\label{eq:totHam}
	H=H_0+\sum_jV_jn_j,
\end{equation}
where only two of the $V_j$ are nonzero. 

Below we shall investigate quench dynamics. We take the initial state at time zero to be the ground state $| \psi(0) \rangle\equiv |\mathcal{E}_0 \rangle_{H}$ of the Hamiltonian \eqref{eq:totHam} with trapping potentials present. At time zero, we turn off the trapping potentials to start the quench dynamics, and after a time $t$, the state has evolved into 
\begin{equation}
    | \psi(t) \rangle=\exp{(-\mathrm{i}H_0 t /\hbar)} | \psi(0) \rangle.
\end{equation}
	
With the purpose of investigating properties of $|\psi(t)\rangle$, we define the density distribution at time $t$ as
\begin{align}
	\rho_j (t) = \langle \psi(t) |n_j |  \psi(t) \rangle - \frac{ M+\nu\sum_kp_k}{N}.
	\label{eq:charge}
\end{align}
The first term $\langle \psi(t) |n_j |  \psi(t) \rangle$ on the right hand side is the expectation value of the number of particles on site $j$ in the state $|\psi(t)\rangle$, and the second term is the average number of particles per site when there are $M+\nu\sum_kp_k$ particles in the system. Note that $\sum_j\rho_j(t)=-\nu\sum_kp_k$, so $\sum_j\rho_j(t)=-1$ for two quasiholes and $\sum_j\rho_j(t)=+1$ for two quasiparticles at half filling. 
 
FQH systems tend to have uniform density in the bulk, and hence nonzero $\rho_j(t)$ can appear due to edge variations or density disturbances due to a potential. To eliminate the edge variations, we shall also sometimes consider the excess particle density
\begin{align}
	\tilde{\rho}_j (t) = \langle \psi(t) |n_j |  \psi(t) \rangle 
	- \langle n_j \rangle_{0},
	\label{eq:charge2}
\end{align}
where $\langle n_j \rangle_{0}$ is the expectation value of $n_j$ computed in the ground state of $H_0$ with $M+\nu\sum_kp_k$ particles.  
	
Our computations involve exact diagonalization. We consider $3$ particles for the case without anyons to ensure that the dimension of the Hilbert space is not too large. When trapping potentials are present, we have\ $M+\nu\sum_kp_k=3$, i.e.,  $M=2$ bosons for the case of positive potentials (two quasiholes), and $M=4$ bosons for the case of negative potentials (two quasiparticles).  We present results for the real-time dynamics of the lattice system with $N=N_x\times N_y=7\times 7$ sites. We have numerically checked that the dynamics is qualitatively the same in other lattices of comparable size, e.g.\ $N=7\times11$, $N=11\times9$, etc.  We concentrate on initial states for which the trapping potentials are located at particular sites. Two cases are considered: case (i), in which the two sites are placed on the middle of the left and right edges, i.e.\ the sites $(1a,4a)$ and $(7a,4a)$ for the $7\times 7$ lattice; case (ii), in which the two sites are moved one site into the bulk from the left and right edges, i.e.\ $(2a,4a)$ and $(6a,4a)$ for the $7\times 7$ lattice.

\section{Quasiholes}\label{sec:hole}

\begin{figure}
	\includegraphics[width=1\linewidth, trim= 50 530 40 20,clip]{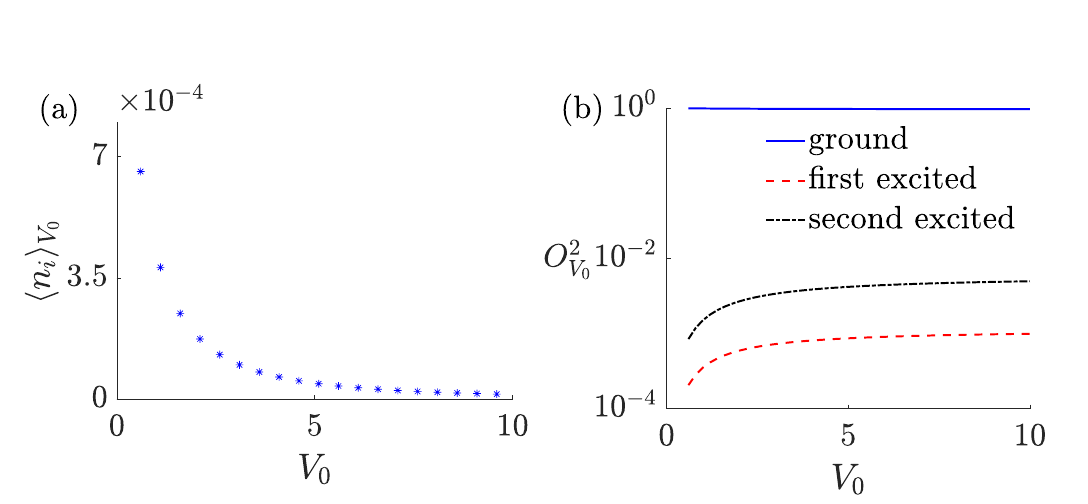}
    \caption{Effects of the pinning potential in the quasihole case with two particles.  (a) Particle density $\langle n_i\rangle_{V_0}$ for the site $(1a,4a)$ (at the edge) versus potential strength $V_0$. (b) Squared overlap $O_{V_0}^2$ between the ground state of the total Hamiltonian $H$ with potentials and the lowest eigenstates of the Hamiltonian $H_0$ without potentials. $V_0$ is in units of hopping strength $J$.} 
	\label{fig:charge_overlap_positive}
\end{figure}
	
We first consider the case of quasiholes, with positive potentials used to pin (localize) them. We choose the following configuration: the number of sites is $N=7\times7$, the number of particles in the system with positive potentials is two, the number of fluxes per plaquette is $\alpha =1/7$, and positive potentials with strength $V_0$ are introduced on the sites $(1a,4a)$ and $(7a,4a)$. In contrast to the case of quasiparticles,  which is shown and discussed in the following section, the chiral motion along the $y$-axis is clearly demonstrated in the case with positive potentials. Moreover, the observed chiral motion is quite robust in the sense that it does not depend on the strength of the potentials or their precise location on the edge.  For example, we obtain similar results if we instead put the potentials on the sites $(2a,4a)$ and $(6a,4a)$. We also find that the chiral motion persists for a very long time.
	
We first investigate the effect of the potential strength $V_0$ on the system by considering the particle density 
\begin{equation}
	\langle n_i \rangle_{V_0} = \langle \psi(0)|n_i|\psi(0) \rangle
\end{equation}
at site $i$ and the overlap 
\begin{equation}
	O_{V_0} = |\langle E_n |{\psi(0) \rangle}|
\end{equation}
between the ground state $|\psi(0)\rangle$ of the total Hamiltonian $H$ with potentials, and the $n$th eigenstate $| E_n \rangle $ of the Hamiltonian $H_0$ with the same number of particles. If $H_0$ has several degenerate eigenstates $|E_n\rangle_m $ at a given energy $E_n$, their total overlap is given by
\begin{equation}
	O_{V_0}=\sqrt{\sum_m|\langle \psi(0) |E_n\rangle_m|^2}.
	\label{eq:overlap_degenerate}
\end{equation}
The squared overlap $O_{V_0}^2$ measures the percentage that the eigenstate $| E_n \rangle $ (or, in case of degeneracy, all the states at energy $E_n$) contributes during the quench dynamics given the initial state is $|\psi(0)\rangle$.
	
In Fig.~\ref{fig:charge_overlap_positive}(a) we plot the particle density $\langle n_i \rangle_{V_0}$ at site $(1a,4a)$. Due to the symmetry under $\pi$ rotations (inversion), the density at this site is equal to the density at site $(7a,4a)$.  We observe that the particle density at site $(1a,4a)$ is quite small even when the potential is not strong, say, $V_0=0.5$. In addition, $\langle n_i \rangle_{V_0}$ drops quickly as $V_0$ increases. This is expected, since the particle density on the site should go to zero when there is a large positive potential on the site.
	
In Fig.~\ref{fig:charge_overlap_positive}(b) we show the squared overlap between the initial state $|\psi(0) \rangle$ and the zeroth, first, and second excited state of $H_0$. The ground state $|E_0 \rangle$ of $H_0$ is non-degenerate, while the first excited state is two-fold degenerate $|E_1 \rangle_{1,2}$. This degeneracy is not lifted even with a very strong potential, say, $V_0=10000$. We can see that the overlap of the ground state dominates over all other overlaps of the excited states. We find that $O_{V_0}^2>0.9604$ for the ground state for all $V_0>0$. We note that this means that there is a large overlap between states with pinned and unpinned quasiholes, as the ground state with two particles has two unpinned quasiholes.
	
\begin{figure}
	\includegraphics[width=1\linewidth, trim= 50 160 50 100,clip]{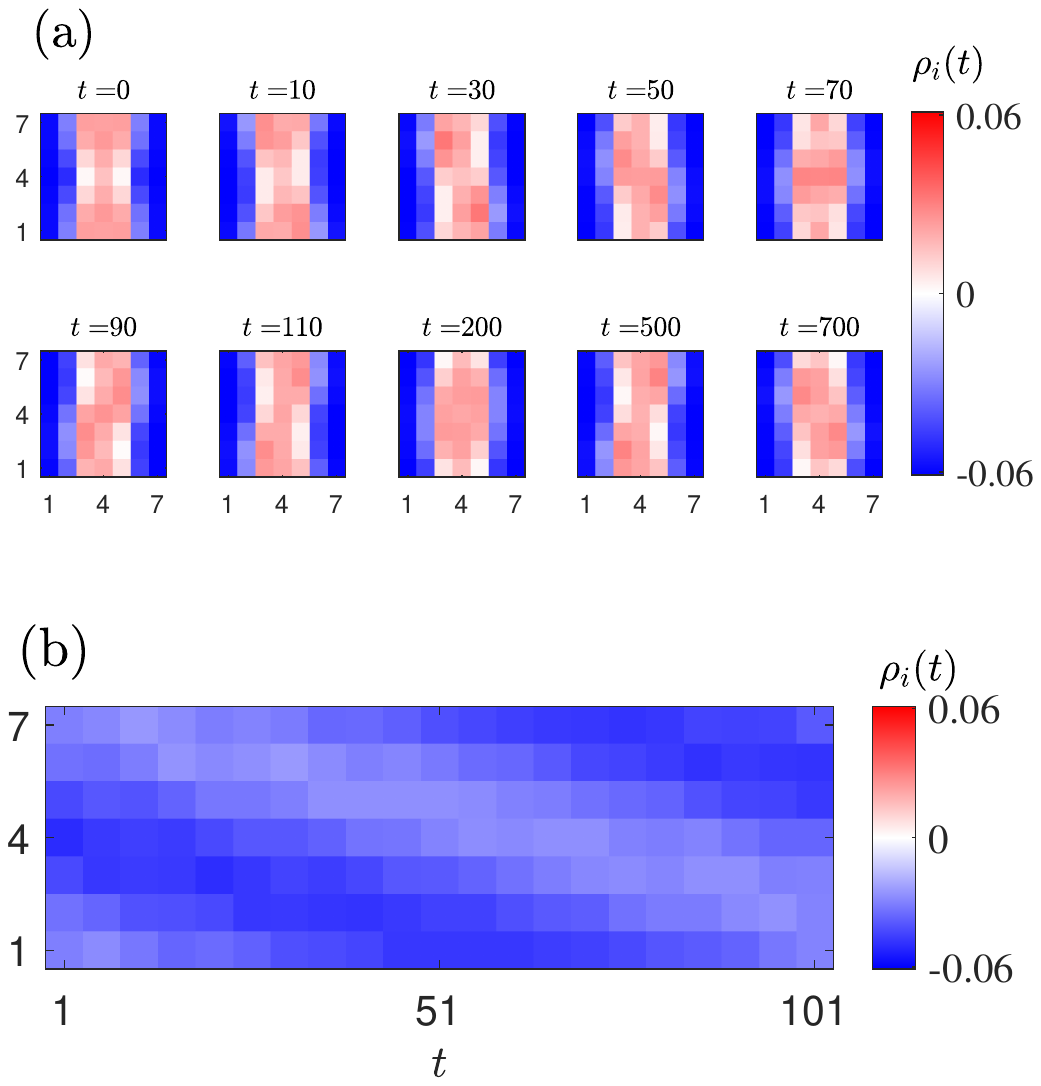}
	\caption{(a) Quench dynamics of the density distribution \eqref{eq:charge} for the quasihole case. The size of the lattice is $7 \times 7$, and the number of particles is two. The trapping potentials are located at the sites $(1a,4a)$ and $(7a,4a)$, i.e.\ at the left and right edges, and have strength $V_0=100$. The flux density is $\alpha=1/7$. (b) The density distribution of the second column of sites, counted from the left, as a function of time for the quench dynamics shown in (a). A staircase pattern is seen, which means that there is chiral motion. Time is in units of inverse of hopping strength, i.e., $\hbar/J$.}
	\label{fig:quench_hole}
\end{figure}
	
In Fig.~\ref{fig:quench_hole} we show the quench dynamics of the system with positive potentials. In Fig.~\ref{fig:quench_hole}(a) we observe that the density distribution moves chirally around the edge (the $y$-axis) in addition to the dispersion along the edge. Meanwhile, there is little density spreading into the bulk sites. This is different from the dynamics observed in the models studied in Refs.~\cite{Nandy2020,dong2018charge} where the particles spread into larger regions in the bulk on longer timescales. In our work, however, the density depletion hardly spreads into the bulk anymore after a short time $t=1$.  The leftmost and rightmost columns (edge columns) do not show much visible variation in density, so it is convenient to use the second column (column next to the edge) to track and visualize the density dynamics.  In Fig.~\ref{fig:quench_hole}(b) we take snapshots of the density distribution of the second column of sites counted from the left, and  concatenate snapshots at different instants to form a spacetime plot.  This visualizes the chiral motion of the density distribution which is seen as a `staircase' pattern in such a plot.  The observed chiral dynamics suggests that we have a quasihole propagating along each of the two edges. 

This picture is supported by monitoring the excess density along the edges.  
We observe that even during long periods of quench dynamics the sum of \eqref{eq:charge2} over the sites along the two leftmost (rightmost) columns of sites equals $1/2$ with a good accuracy, $\tilde{\rho}_{\mathrm{edge}}=\sum_{i:x_i\leq 2} \tilde{\rho}_i(t) \approx -0.49 $. 
The total excess density corresponding to half a particle near each edge is consistent with the expectation that quasiholes are created and trapped.  We note however that the system is inversion-symmetric, and there is one particle missing compared to the case without potentials, so a similar density pattern could in principle also arise in non-topological systems. The interpretation of the density depletions as quasiholes is further corroborated by the analysis of model wavefunctions in Sec.~\ref{sec:modelstate}. The nearly-constant value of $\tilde{\rho}_{\mathrm{edge}}$ over time suggests that the quasiholes stay confined to the edges as they evolve in time.

We find similar results as above when we examine the quench dynamics for various sizes of the lattice, different locations of the trapping potentials (at the edges or one site away from the edges) and various potential strengths $V_0$. In fact, we find that $\tilde{\rho}_{\mathrm{edge}}\approx -0.5$ even if no potentials are present ($V_0=0$),  although in such a case the excess density is spread uniformly in the $y$ direction and hence chiral motion is not visible. But as soon as the potentials are strong enough to create a discernible ``bulge'' of the excess charge near their locations, the chiral motion can be observed. We verified that this is the case even for weak potentials, such as $V_0=0.5$.

Moreover, the chiral motion is also observed for values of the flux density, say, $\alpha=6/49$, which deviate slightly from the correct value $\alpha=1/7$. (In fact, if we replace $N_{\textrm{plaq}}$ with $N$ in Eq.~\eqref{eq:filling_plaq} we get the filling factor $1/2$ for $\alpha=6/49$. The ambiguity in defining the magnetic flux through the lattice, when edges are present, is thus not a severe problem for the considered systems.) These results indicate that the chiral dynamics of quasiholes pinned by the positive potentials is quite robust in small lattices. In addition, we observe that the chiral motion of the quasiholes persists for a large timescale $t>700$.

\begin{figure}
	\includegraphics[width=1\linewidth, trim= 50 150 30 100,clip]{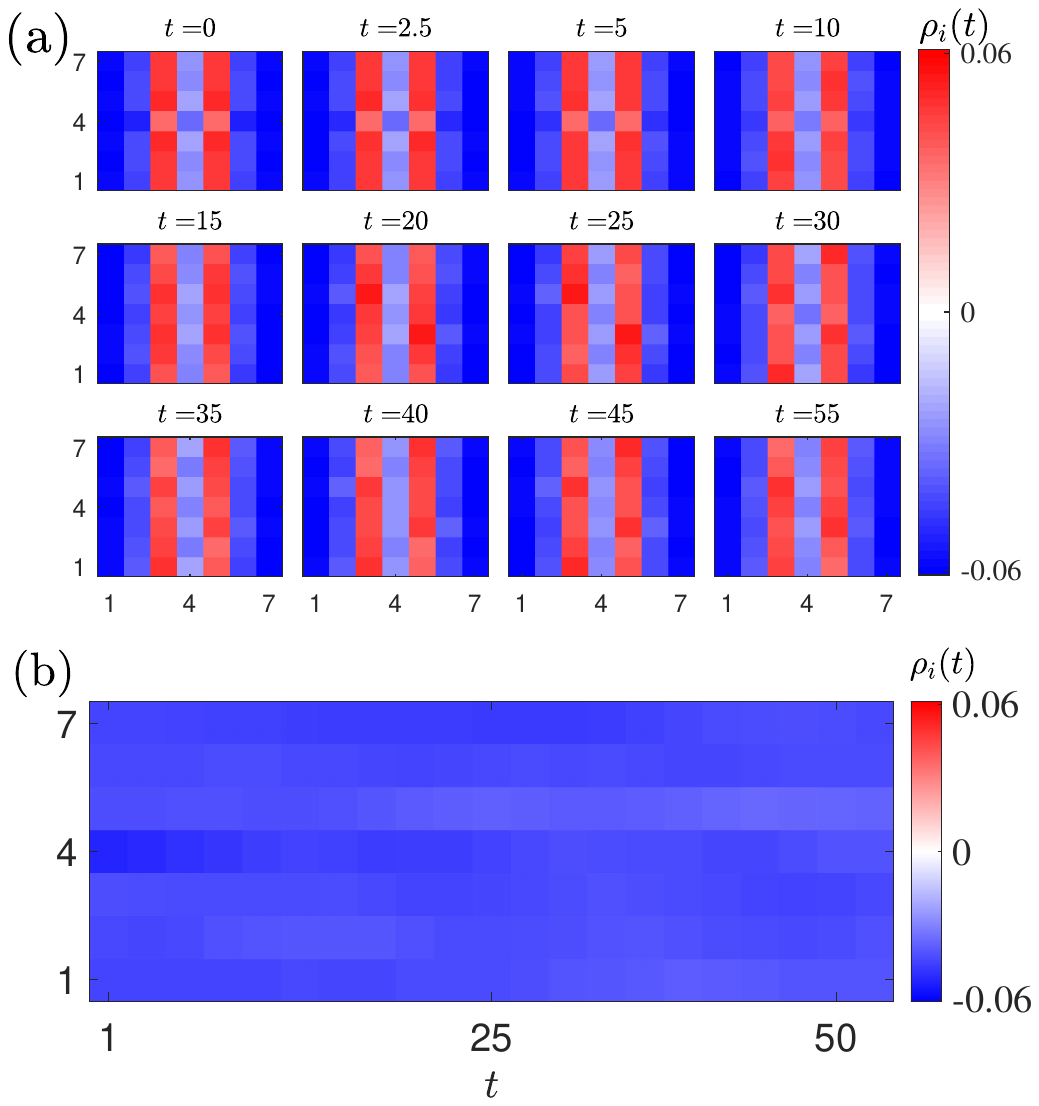}
	\caption{(a) Snapshots of the time-evolving density distribution \eqref{eq:charge} for the system with quasiholes and a flux density $\alpha=3/10$, significantly different from the `correct' value $\alpha=1/7$. The trapping potentials ($V_0=100$) are at the edges, $(1a,4a)$ and $(7a,4a)$. (b) The density distribution of the second column of sites, counted from the left, as a function of time for the quench dynamics shown in (a). The lack of a clear staircase pattern signals the absence of chiral motion. Time is in units of inverse of hopping strength, i.e., $\hbar/J$.}
	\label{fig:quench_hole_6/20}
\end{figure}

If the flux density deviates much from the correct value $\alpha=1/7$, one would expect that the chiral motion is absent. In Fig.~\ref{fig:quench_hole_6/20}, we show the quench dynamics of the system with the same configurations in Fig.~\ref{fig:quench_hole}, but with the flux density $\alpha=3/10$. As expected, no chiral motion is observed.

\begin{figure}
	\includegraphics[width=1\linewidth, trim= 70 150 30 100, clip]{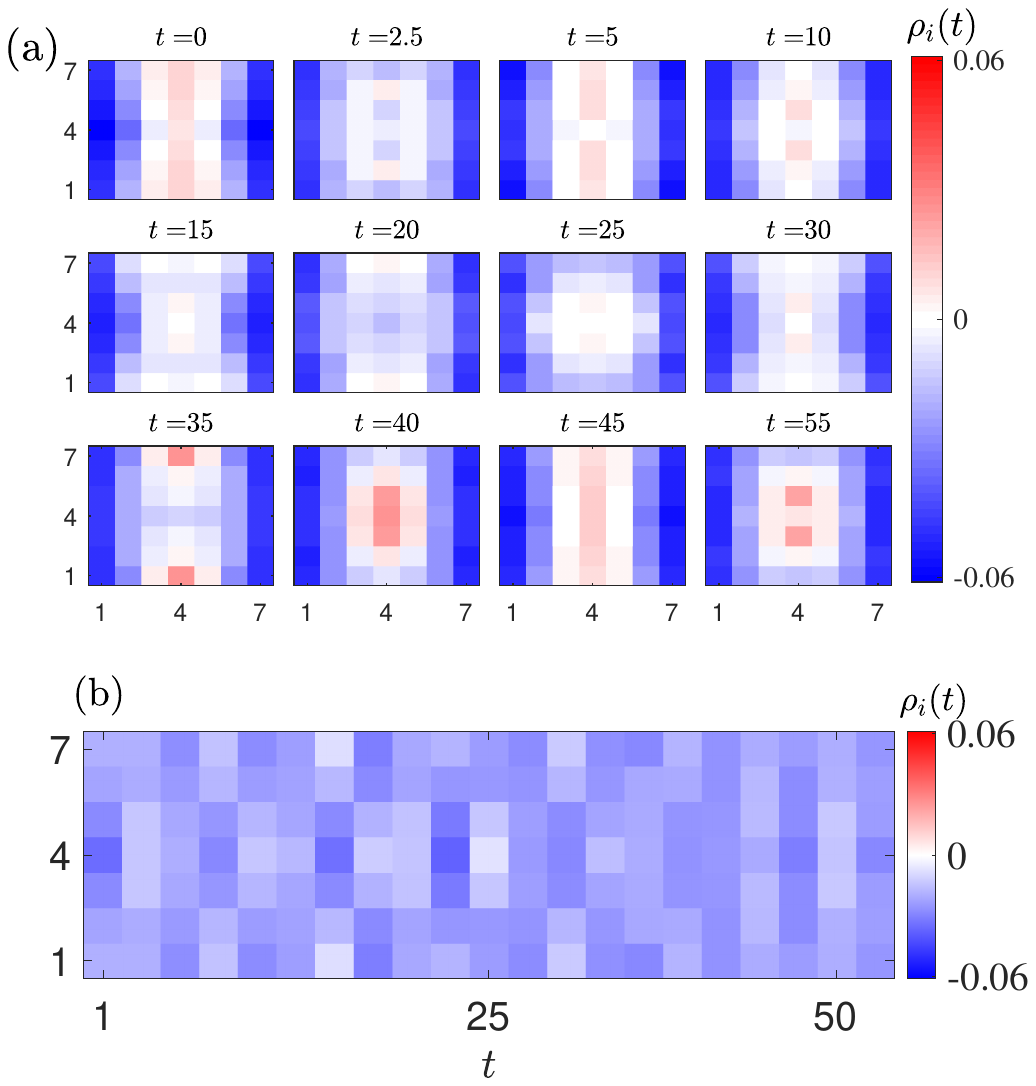}
	\caption{(a) Snapshots of the density distribution \eqref{eq:charge} for the quasihole case in the absence of the magnetic field, i.e.\ $\alpha=0$. The trapping potentials ($V_0=100$) are located at the edge sites $(1a,4a)$ and $(7a,4a)$. (b) The density distribution of the second column of sites, counted from the left, as a function of time for the quench dynamics shown in (a). When the magnetic field is turned off, the model has mirror symmetry both between up and down and between left and right, and hence there is no chiral motion. Time is in units of inverse of hopping strength, i.e., $\hbar/J$.}
	\label{fig:quench_hole_0}
\end{figure}

Furthermore, we consider an extreme case in which the magnetic field is turned off, $\alpha=0$. The result is shown in Fig.~\ref{fig:quench_hole_0}. In this case, there is no chiral motion. Instead, we observe that the density distribution at the sites of the left (and right) half of the lattices split into two parts with equal weight, and two regions of depleted density move with opposite chirality (upward and downward directions).

\section{Quasiparticles}\label{sec:electron}

In this section, we consider the case with four particles.  We use negative potentials for creating the initial state.  In this way we attempt to  spatially localize quasiparticles. We find that the trapped objects have total excess particle density close to 0.5, as quasiparticles are expected to, but that they do not display clear chiral dynamics.
	
\begin{figure}
	\includegraphics[width=1\linewidth, trim= 50 550 50 20,clip]{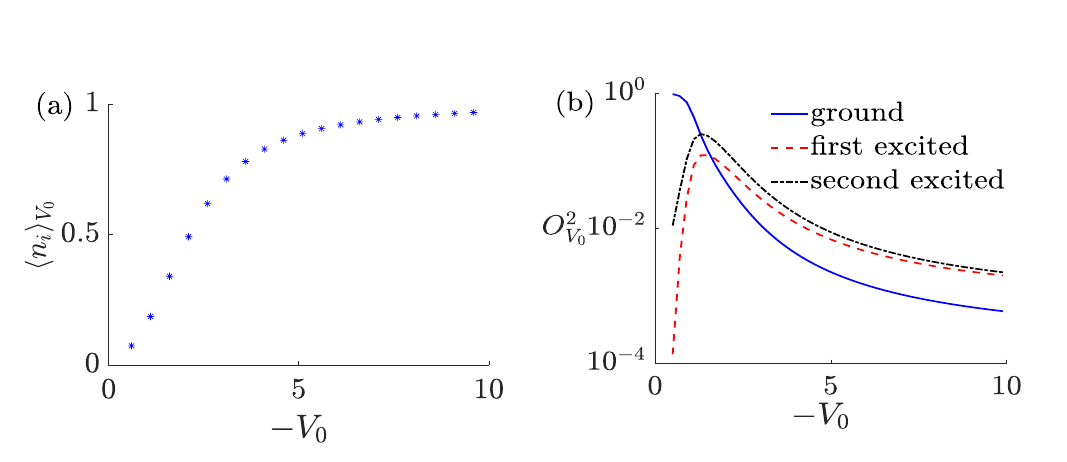}
	\caption{Effects of pinning potential in the quasiparticle case with four particles.  (a) Particle density $\langle n_i\rangle_{V_0}$ at site $(1a,4a)$ versus potential strength $-V_0$. (b) Squared overlap $O_{V_0}^2$ between the ground state of the total Hamiltonian $H$ with potentials and the lowest eigenstates of the Hamiltonian $H_0$ without potentials. $V_0$ is in units of hopping strength $J$.}
	\label{fig:charge_overlap_negative}
\end{figure}

In Fig.~\ref{fig:charge_overlap_negative}(a), we show the particle density at site $(1a,4a)$ as a function of the potential strength $-V_0$.  As $-V_0$ increases, $\langle n_i\rangle_{V_0}$ grows and eventually converges to its maximum possible value, which is $1$. This is expected because the stronger the on-site negative potential is, the more the energy is reduced by having a particle on that site. We avoid potentials close to zero as $|\psi(0)\rangle$ has a degeneracy for $V_0=0$. 
	
\begin{figure}
	\includegraphics[width=1\linewidth, trim= 50 250 50 20,clip]{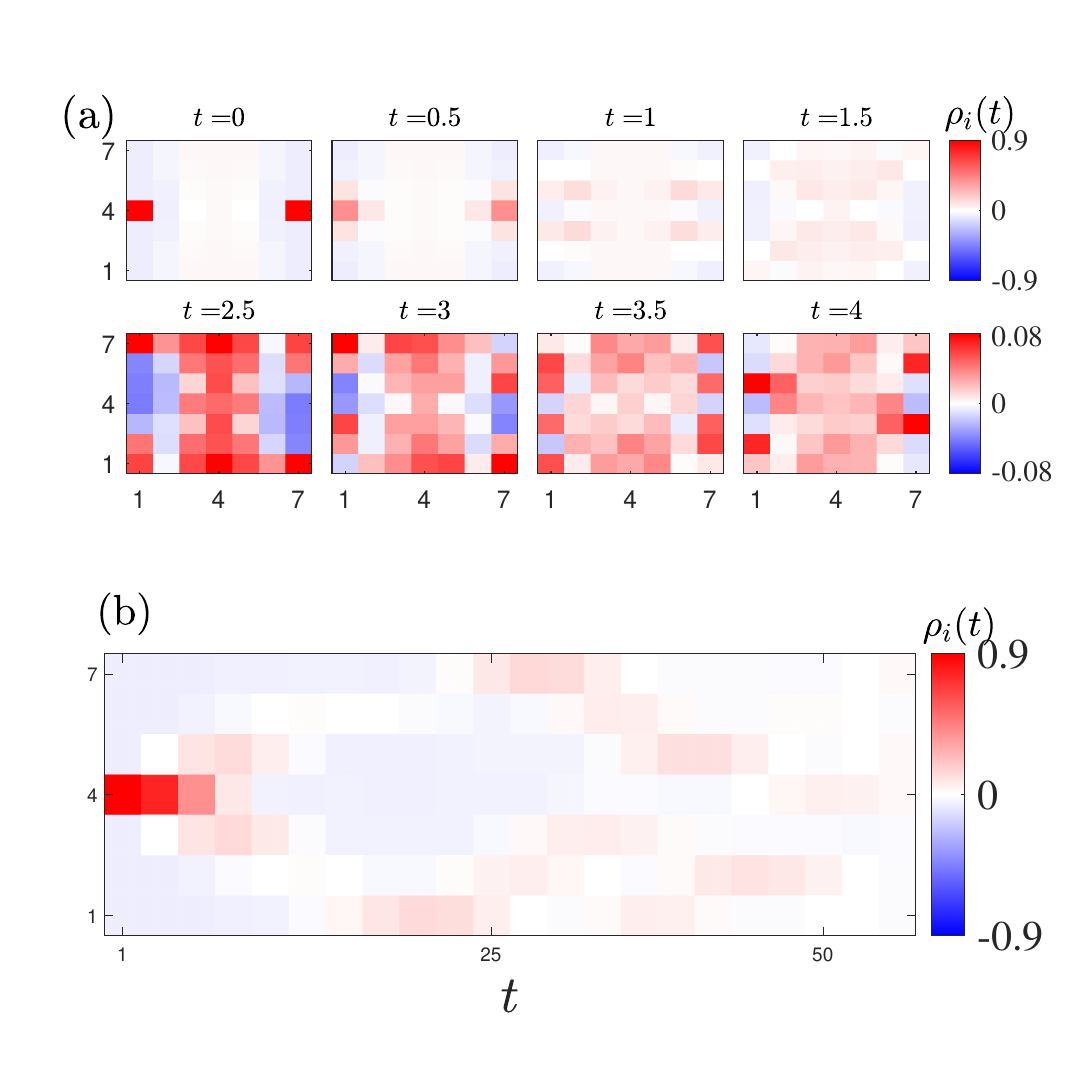}
	\caption{(a) Snapshots of the density distribution \eqref{eq:charge} in the quasiparticle case.  The trapping potentials (strength $V_0=-100$) are at the edge sites $(1a,4a)$ and $(7a,4a)$. A different color scale has been used in this plot, because the particles spread into the bulk sites on a longer timescale. (b) The density distribution of the leftmost column of sites as a function of time for the quench dynamics shown in (a). No staircase pattern is seen, which signals the absence of chiral motion. Time is in units of inverse of hopping strength, i.e., $\hbar/J$.}
	\label{fig:quench_electron}
\end{figure}

In Fig.~\ref{fig:charge_overlap_negative}(b) we show the squared overlap $O_{V_0}^2$ between the initial state $|\psi(0) \rangle$ and the lowest eigenstates of $H_0$ with four particles as a function of $|V_0|= -V_0$.  The ground states of $H_0$ are doubly degenerate, and thus the overlap is calculated using \eqref{eq:overlap_degenerate}.  As $-V_0$ increases, more and more higher excited states have appreciable overlap and thus  contribute to the quench dynamics. We observe that the overlaps with respect to $|E_n \rangle$ depend substantially on $|V_0|$, and many of them have comparable magnitude.  The total overlap with respect to the doubly degenerate ground state $|E_0 \rangle_{1,2}$ decreases rapidly with  $-V_0$; its magnitude does not dominate over other eigenstates for $-V_0\gtrsim 1.5$. This behavior is very different from the case of  quasiholes trapped by positive potentials, shown in Fig.~\ref{fig:charge_overlap_positive}(b).  In that case the ground state dominates even for large potentials. 
	
In Fig.~\ref{fig:quench_electron} we demonstrate the quench dynamics for negative potentials located at the edges with strength $V_0=-100$ which is huge compared to the hopping strength. In Fig.~\ref{fig:quench_electron}(a) we show the dynamics of the density distribution, and in Fig.~\ref{fig:quench_electron}(b) we plot the density distribution of the first column of sites, counted from the left, as a function of time. The particle density \eqref{eq:charge2} in the initial state has an excess density of $\tilde{\rho}_{\mathrm{edge}}(0)=\sum_{i:x_i\leq 2} \tilde{\rho}_i(0) \approx 0.51 $ in the two leftmost (two rightmost) columns of sites. That is, at the beginning of the dynamics we have an excess density of approximately a half of a particle (this time with opposite sign) located near each edge, consistent with the expectation that we trapped quasiparticles. Again we note that the system has inversion symmetry, and there is one particle extra compared to the case without potentials, so a similar excess density could also appear without the presence of topology. Further support that quasiparticles are trapped is presented in Sec.~\ref{sec:modelstate}.
	
A significant difference from the results in Sec.~\ref{sec:hole} appears when the density evolves in time. From Figs.~\ref{fig:quench_electron}(a) and \ref{fig:quench_electron}(b), we observe no chiral motion of the density distribution, but patterns that are different from the case of quasiholes. We find that, during quench dynamics the excess density roughly splits into two parts and each of them move in opposite directions. Meanwhile, in contrast to the quasihole case, the excess density spreads into the bulk with much weight in the beginning, but stops spreading more into the bulk sites after some time. 
	
In addition, we have varied $V_0$ and also varied the positions of the trapping potentials, e.g., at the edges, or one or two sites away from the edge.  We have not observed clear chiral motion in any of these cases.  Furthermore, we find that the patterns of quench dynamics for negative potentials vary substantially as we change the strength and/or locations of the pinning potentials, which was not the case for quasiholes trapped by positive potentials.  We also tried to use pinning potentials over a block of sites to create quasiparticles as in \cite{Raciunas2018}.  Also in that case, we have not seen clear chiral motion.

\section{Model quasiparticle state}\label{sec:modelstate}
	
Could the difference between the dynamics for positive and negative potentials arise because the former succeeds in pinning anyons, and the latter does not? To further study the question whether the trapped objects are quasiholes and quasiparticles, we here discuss a model wavefunction with anyonic excitations \cite{tu2014lattice,glasser2016lattice,nielsen2018quasielectrons}.
	
We write the position of site $j$ as a complex number $\xi_j=x_j+iy_j$. Then, we define a mapping from a cylinder to a complex plane: $z_j=\exp(2\pi\xi_j/L_y)$, where $L_y$ is the circumference of the cylinder (for a square lattice with unit lattice constant we have $L_y=N_y$). For brevity, we write all $z_j$s as a vector $\mathbf{z}=[z_1,z_2,\dots, z_N]$. Like any state of a hardcore boson system, the model lattice Laughlin state $|\psi_{\mathrm{model}}\rangle$ can be written in the occupation number basis,
\begin{equation}
	|\psi_{\mathrm{model}}\rangle=\frac{1}{C}\sum_{\mathbf{n}}\psi_{\mathrm{model}}(\mathbf{z},\mathbf{n})|\mathbf{n}\rangle, 
	\label{eq:occbasis}
\end{equation}
where $\mathbf{n}=[n_1,n_2,\dots, n_N]$ is the vector of the occupation numbers of each site, with $n_j\in\{0,1\}$, $|\mathbf{n}\rangle$ is the corresponding basis state, $\psi_{\mathrm{model}}(\mathbf{z},\mathbf{n})$ are the unnormalized wavefunction coefficients and $C$ is the normalization constant. For a model lattice Laughlin state without anyons on a cylinder \cite{glasser2016lattice}, the coefficients are
\begin{multline}
	\psi_{\mathrm{model}}(\mathbf{z},\mathbf{n})=
	\delta(qM+p_{+}+p_{-}-N\eta)\prod_{j}\chi_j^{n_j}\times\\ \times\prod_{j}z_j^{p_{-}n_j}
	\prod_{j<k}(z_j-z_k)^{\gamma_j\gamma_k}.
	\label{eq:laughlin_cylinder}
\end{multline}
Here, $\delta(qM+p_{+}+p_{-}-N\eta)$ is the Kronecker delta fixing the charge neutrality, i.e.\ the number of particles $M$ (with $M=\sum_jn_j$) and $\gamma_j=(q n_j-\eta)/\sqrt{q}$, with $\eta$ being the flux assigned to each site (in our case, we put $\eta=\alpha$). The integer $q$ determines the topological order of the wavefunction and in our case $q=2$. The gauge factors $\chi_j$ are arbitrary complex numbers with $|\chi_j|=1$. We note that in this work we use the word ``gauge'' in two meanings: the Landau gauge \eqref{eq:landaugauge} of the hoppings of the Hofstadter model \eqref{eq:BH_1}, and the factors $\chi_j$ of the model wavefunctions. To avoid confusion, we will always refer to the latter as ``the gauge factors $\chi_j$''. 
	
The real numbers $p_{-}$, $p_{+}$, have to fulfill the relation
\begin{equation}
	p_+=p_--q+2\eta.
	\label{eq:ppluspminus}
\end{equation}
We note that the formulation in \cite{tu2014lattice,glasser2016lattice} of the lattice Laughlin states on a cylinder did not contain $p_-$ and $p_+$. The origin and meaning of these parameters is described in Appendix \ref{app:inversion}, while the relation of \eqref{eq:laughlin_cylinder} to the continuum states is described in Appendix \ref{app:LatticeVsContinuum}. 
	
In our calculations, we want to use a state with a given $M$, $N$ and $\eta=\alpha$. Thus, we determine $p_-$ from a combination of the charge neutrality condition and \eqref{eq:ppluspminus},
\begin{equation}
	p_{-}=\frac{1}{2}\left(\left(N-2\right)\eta-q\left(M-1\right)\right),
	\label{eq:condition3}
\end{equation}
and then $p_+$ from \eqref{eq:ppluspminus}.
	
Some properties of \eqref{eq:laughlin_cylinder}, such as the particle density and the topological order, are  independent from the gauge factors $\chi_j$. Its overlap with the ground state of the Hofstadter model, however, strongly depends on $\chi_j$. Therefore, we optimize these factors numerically to maximize the overlap. For the $7\times 7$ system with $M=3$ and $\alpha=1/7$, we obtain the squared overlap $|\langle \psi(0) |\psi_{\mathrm{model}} \rangle |^2\approx 0.984$ after optimization.
	
We can add anyonic excitations to the state \eqref{eq:laughlin_cylinder}. They are characterized by the integers $p_k$ (see Eq.~\eqref{eq:filling_plaq}), with $p_k=1$ and $p_k=-1$ corresponding to a basic quasihole and quasiparticle, respectively, and the complex positions $\omega_k\in\mathbb{C}$. Note that the complex positions are not required to coincide with lattice sites \cite{nielsen2018quasielectrons}. After mapping to the complex plane, these locations transform into $w_k=\exp(2\pi\omega_k/L_y)$. The wavefunction with anyons is then
\begin{multline}
	\psi_{\mathrm{model}}(\mathbf{z},\mathbf{n},\mathbf{w},\mathbf{p})=\\
	\delta(qM+p_{+}+p_{-}+\sum_{k}p_k-N\eta)
	\prod_{j}\chi_j^{n_j}\prod_{j}z_j^{p_{-}n_j} \times \\
	\times
	\prod_{j,k}(z_j-w_k)^{p_kn_j}\prod_{j<k}(z_j-z_k)^{\gamma_j\gamma_k}.
	\label{eq:laughlin_cylinder_anyons}
\end{multline}
The $w_k$ have the braiding properties expected for anyons in systems with Laughlin type topology when the $w_k$ remain well-separated during the entire braiding process \cite{nielsen2018quasielectrons}. 
This is true even if $p_k$ is negative -- while such a wavefunction does not have a consistent continuum limit \cite{patra2021continuum}, it is a valid quasiparticle ansatz on lattices with finite lattice constant, yielding correct quasiparticle charge and statistics \cite{nielsen2018quasielectrons}. On small lattices with low densities, the changes in the particle density caused by the anyons may not be fully separated.  

We consider two examples in a $7\times 7$ system with $\eta=\alpha=1/7$: two quasiholes ($M=2$) and two quasiparticles ($M=4$). In both cases, we place the anyons at the same positions as the pinning potentials in Secs.\ \ref{sec:hole} and \ref{sec:electron} (plus a small shift to avoid infinities in \eqref{eq:laughlin_cylinder_anyons}). We use the same $\chi_j$ as in the $M=3$ case (i.e.\ we do not re-optimize them). The resulting squared overlaps with ground states of the Hofstadter model with potentials ($V_0=\pm 100$) are $|\langle \psi(0) |\psi_{\mathrm{model}} \rangle |^2 \approx 0.976$ and $|\langle \psi(0) |\psi_{\mathrm{model}} \rangle |^2 \approx  0.977$ for quasiholes and quasiparticles, respectively.
	
Therefore, the ground states of the Hofstadter model states are similar to the model states, but not perfectly equal to them. To check whether the discrepancy is the reason for the lack of chiral dynamics in the quasiparticle case, we also compute the quench dynamics generated by the hopping Hamiltonian $H_0$ (Eq.~(\ref{eq:BH_1})) with the model state \eqref{eq:laughlin_cylinder_anyons} with two quasiparticles as the initial state. The results are quite similar to Fig.~\ref{fig:quench_electron}, i.e.\ we do not observe clear chiral dynamics.
	
\begin{figure}
	\includegraphics[width=0.25\textwidth]{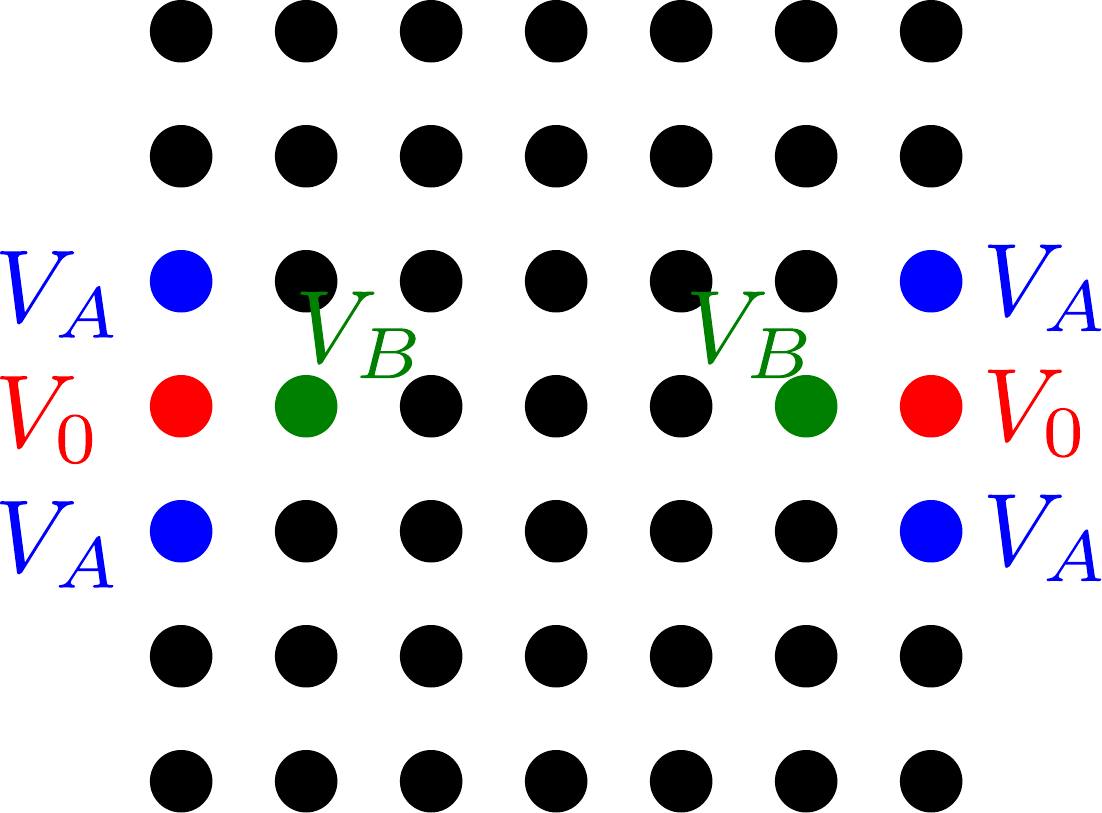}
	\caption{The locations and strengths of the potentials that we optimize to increase the overlap of the ground state of the Hamiltonian with negative $V_0$ with the model wavefunction \eqref{eq:laughlin_cylinder_anyons} with two quasiparticles. During the optimization, $V_0=-100$ stays constant, while $V_A$ and $V_B$ are adjusted. $V_0$ is in units of hopping strength $J$.}
	\label{fig:VAVB}
\end{figure}
	
We note that we can increase the overlap between the quasiparticle state of the Hofstadter model and the model lattice Laughlin quasiparticle state by introducing auxiliary potentials $V_A$ and $V_B$ at the nearest-neighboring sites of the original potentials (see Fig.\ \ref{fig:VAVB}), and optimize them to maximize the overlap. In such a way, we obtained the squared overlap $|\langle \psi(0) |\psi_{\mathrm{model}} \rangle |^2=0.993$. However, using the ground state with the optimized potentials as the initial state also does not lead to chiral dynamics.
	
The results presented in this section suggest that the difference between the cases of positive and negative potentials is not related to the ability to pin anyons, as both quasiholes and quasiparticles seem to be successfully created and pinned by the pinning potentials.

\section{Discussion and Conclusions}\label{sec:conclusions}

We have studied quench dynamics in  small lattices with cylindrical boundary conditions, using interacting (hardcore) bosons subjected to the Hofstadter Hamiltonian. We considered particle numbers appropriate to both two quasiholes ($M=2$) and two quasiparticles ($M=4$), relative to a FQH state with filling factor $\nu=1/2$.  In the two cases we used positive and negative pinning potentials so that the initial positions of the quasiholes and quasiparticles are localized near the edges of the lattices.  The quench dynamics begins when the trapping potentials are turned off.  One expects to observe chiral motion of the excess density due to the presence of chiral edge modes.  Our  results for the two cases are very different. While it is generally easy to observe chiral motion of quasiholes (negative excess density), and the observed chiral dynamics are quite robust, we have not found clear chiral motion of quasiparticles (positive excess density). 
	
The absence of the chiral motion of quasiparticles was further verified by investigating the quench dynamics with the model lattice Laughlin state as the initial state. Moreover we introduced auxiliary potentials in the Hofstadter model to optimize the overlap between the ground state of the Hofstadter Hamiltonian and the model lattice Laughlin state. The chiral dynamics was still absent.

Why is there such a pronounced difference between the quasihole and quasiparticle cases? Due to the small particle densities, the details of pinning are quite different in the two cases---locally decreasing the  density is a very different process compared to locally increasing the density. In the former case (quasiholes, negative excess density) there is a natural limit to how far the local density $\langle n_i\rangle_{V_0}$ can be reduced, since $\langle n_i\rangle_{V_0}$ is small already for $V_0=0$. In the latter case (quasiparticles, positive excess density), the site density can increase enormously, all the way up to $1$, leading to the excitation of many high-energy modes whose nature might be unrelated to FQH physics. Notice that this is true with the hardcore constraint $U/J \rightarrow \infty$. These high-energy contributions can be expected to mask or disturb the topological and chiral physics that FQH quasiparticles should show. This difference between the two cases is mirrored in the overlaps shown in Figs.~\ref{fig:charge_overlap_positive}(b) and \ref{fig:charge_overlap_negative}(b).  

This difference raises the possibility that the pinning of quasiparticles might be disrupted, unlike the pinning of quasiholes.  However, the comparison with model quasiparticle states (Sec.\ \ref{sec:modelstate}) indicates that both positive and negative potentials pin the anyons successfully.  We therefore conclude that the difference must come from the contributions of excited states in the dynamics --- the quasihole dynamics is dominated by the ground state, while the quasiparticle dynamics involves a number of highly excited states, as indicated in Figs.~\ref{fig:charge_overlap_positive}(b) and \ref{fig:charge_overlap_negative}(b).  The disruption of chiral dynamics is a \emph{dynamical} effect. 

The absence of chiral motion suggests that the dynamics of quasiparticles in small lattices with low particle densities, of the sizes considered in this work, is different from that in large lattices, and from that in two-dimensional electron gases.  This has important implications for the study of FQH physics in novel platforms such as that of Ref.\ \cite{leonard2022}, where the particle density and system size might be naturally small.  Our results indicate that, in such situations, quasiparticle dynamics is fragile as there is a natural  tendency to create substantial additional (possibly unwanted) excitations when localized quasiparticles are created.  Remarkably, the difficulty with observing quasiparticle features shows up primarily in the dynamics, and not necessarily in creating or pinning them.  

This work opens up a number of questions:  (1) Our conjectured explanation above suggests that quasiparticles and quasiholes would behave more similarly if the average particle density (lattice filling) were close to $1/2$.  It would be worthwhile to check this explicitly; however in our setup this is computationally too challenging. Systems with larger numbers of sites and particles can be studied using tensor network methods~\cite{Pollmann2013}. (2)  A detailed understanding of how the dynamical behaviors change with system size (e.g., with constant density or with constant particle number) is currently lacking.  An intriguing question is whether chiral motion of quasiparticles become readily observable for large lattice sizes even when the particle density remains small. (3) It would also be interesting to investigate how the presence or absence of inversion symmetry affects the excess density and dynamics in small systems. (4) We here considered the case of Abelian anyons. In future work it would be interesting to  study the quench dynamics of FQH systems which host non-Abelian anyons.

\begin{acknowledgments}
This work has been supported by Danmarks Frie Forskningsfond under grant number 8049-00074B. X.L. acknowledges the support by The Key Research Foundation  of Education Ministry of Anhui Province (2023AH050073). The research of M.H. is supported by the Deutsche Forschungs gemeinschaft through SFB No. 1143 (Project ID No.
597 247310070).
\end{acknowledgments}

\appendix

\section{Inversion-symmetric lattice Laughlin wavefunctions on a cylinder}\label{app:inversion}
	
In this Appendix, we derive the expression \eqref{eq:laughlin_cylinder} and condition \eqref{eq:ppluspminus} from the known expressions for lattice Laughlin wavefunctions \cite{tu2014lattice,glasser2016lattice} by demanding inversion symmetry. 
 
Let us start by considering a planar system of $N$ sites at locations described by complex numbers $z_j=x_j+iy_j$, with $j=1,2,\dots{N}$, populated by $M$ particles: either by hardcore bosons or fermions. To each site, we assign a flux $\eta\in \mathbb{R}$ in units of the flux quantum, so the total number of flux quanta in the system is $N\eta$. We also place a compensating charge of $\eta/q$ (in units where a single particle has charge $-1$) at each site, in analogy to a continuum Landau level where a uniform background charge is used.
	
According to \cite{tu2014lattice,glasser2016lattice}, the coefficients of a lattice Laughlin wavefunction on a plane (see Eq.~\eqref{eq:occbasis}) are
\begin{multline}
	\psi_{\mathrm{model}}(\mathbf{z}, \mathbf{n})=\\
	\delta(qM-N\eta)\prod_{j}\chi_j^{n_j} \prod_{j<k}(z_j-z_k)^{\gamma_j\gamma_k},
	\label{eq:laughlin}
\end{multline}
(see below Eq.\ \eqref{eq:laughlin_cylinder} for an explanation of the notation). One can add one or more anyonic excitations to \eqref{eq:laughlin}, \cite{glasser2016lattice,nielsen2018quasielectrons} which transforms it into
\begin{multline}
	\psi_{\mathrm{model}}(\mathbf{z}, \mathbf{n}, \mathbf{w}, \mathbf{p})=\\=
	\delta(qM+\sum_{k}p_k-N\eta)\prod_{j}\chi_j^{n_j}\prod_{j,k}(z_j-w_k)^{p_kn_j}\times \\
	\times \prod_{j<k}(z_j-z_k)^{\gamma_j\gamma_k},
	\label{eq:laughlin_anyons}
\end{multline}
where, similarly to \eqref{eq:laughlin_cylinder_anyons}, $p_k\in\mathbb{Z}$ and $p_k/q$ is the charge of the $k$th anyon. The coordinate $w_k\in\mathbb{C}$ is its position. Both $p_k$ and $w_k$ are external parameters of the wavefunctions. Note that $w_k$ is not required to be equal to one of the $z_j$, but can take any value in the complex plane.
	
Equations \eqref{eq:laughlin} and \eqref{eq:laughlin_anyons} can be generalized to a cylinder \cite{tu2014lattice,glasser2016lattice}. Let us consider a plane of size $L_x\times L_y$, whose upper and lower edges are glued to make a cylinder (i.e.\ $y$ is the periodic direction, and $L_y$ is the circumference of the cylinder). If $\xi_j$ and $\omega_i$ are the coordinates of sites and anyons, respectively, within the $L_x\times L_y$ rectangle, then the substitution $z_j=\exp(2\pi\xi_j/L_y)$, $w_j=\exp(2\pi\omega_j/L_y)$ turns \eqref{eq:laughlin} and \eqref{eq:laughlin_anyons} into cylinder wavefunctions \cite{tu2014lattice}. 
	
\begin{figure}
	\includegraphics[width=0.5\textwidth]{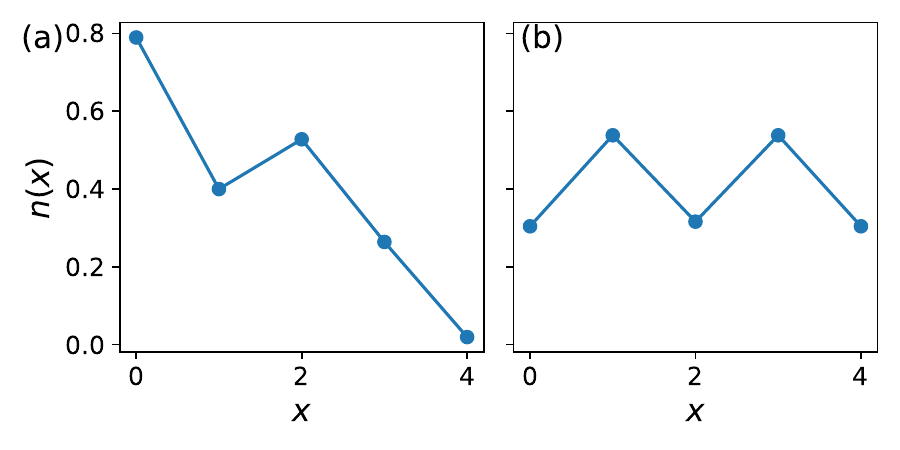}
	\caption{(a) The average particle number as a function of the (non-periodic) coordinate $x$ for a $5\times 5$ square lattice, for the ``naive'' implementation of a $q=2$, $M=2$ lattice Laughlin wavefunction on a cylinder, i.e.\ Eq.~\eqref{eq:laughlin} with $z_j=\exp(2\pi\xi_j/L_y)$ and $\eta=4/25$. (b) The same quantity for the modified wavefunction \eqref{eq:laughlin_cylinder} with $p_-$ given by \eqref{eq:condition3}, and $q$, $M$ and $\eta$ the same as in (a).}
	\label{fig:CompareLatticeFigure}
\end{figure}

It was shown that these cylinder wavefunctions have Laughlin-type topological order \cite{tu2014lattice,glasser2016lattice}. However, while the particle density in the continuum cylinder wavefunction with no anyons seems to be invariant under inversion \cite{rezayi1994laughlin}, in the lattice cylinder wavefunction it is not (except from the special case $\eta=q/2$). An example for a $5\times 5$ square lattice with $q=2$ and $M=2$ is shown in Fig.\ \ref{fig:CompareLatticeFigure}(a). The figure shows the average particle number at a given $x$ coordinate, defined as $n(x)=\sum_j\delta(x-x_j)\langle \psi_{\mathrm{model}} | n_j | \psi_{\mathrm{model}} \rangle$.
	
The lack of inversion symmetry is alarming for two reasons. First, in real quantum Hall systems exchanging the left and right side of the cylinder should not matter. Second, on the plane, the lattice wavefunctions approach the continuum wavefunction in the limit of $\eta\rightarrow 0$ (i.e.\ infinite number of sites per flux quantum) \cite{tu2014lattice}. Similar arguments should apply for a cylinder \cite{glasser2016lattice}, and if this is indeed the case, we should explain why the inversion invariance does not exist for general $\eta$ and is restored at $\eta=0$.
	
In the following, we will show that the inversion invariance of \eqref{eq:laughlin} on a cylinder can be restored at any $\eta$ by choosing the numbering of the sites appropriately and adding appropriate charges, described by the numbers $p_{-}, p_{+} \in \mathbb{R}$, at the coordinates $w_-=0$ and $w_+=\infty$, respectively. They should be understood as infinitely thin, immobile flux tubes. If we plot the complex numbers $z_j$ on the plane, they form rings of different radii. The flux tube at ($w_-=0$) is at the center of the rings, while the flux tube at ($w_+=\infty$) is located outside the rings, infinitely far away. As a result, the effects of the charge $p_+$ appears only in the charge neutrality condition \cite{glasser2016lattice}. The wavefunction is thus given by \eqref{eq:laughlin_cylinder} (but so far, we have not determined the values of $p_-$ and $p_+$).  
	
In our considerations, we will permute and modify the $z_j$ and $n_j$. Therefore, we will write the individual arguments explicitly, i.e.\
\begin{multline}
	\psi_{\mathrm{model}}(\mathbf{z},\mathbf{n})=\\
	\psi_{\mathrm{model}}(z_1, z_2, \dots, z_N; n_1, n_2, \dots, n_N).
\end{multline}
Inversion of the coordinates $\xi_j$ around the point $\xi_c$ transforms $\xi_j-\xi_c\to-(\xi_j-\xi_c)$ and hence 
\begin{equation}
	z_j\rightarrow c/z_j, 
	\label{eq:inversion}
\end{equation}
where $c=e^{4\pi\xi_c/L_y}$. For now, we will consider a general lattice, which is not necessarily inversion invariant. As an intermediate step, we will investigate when
\begin{multline}
	\psi_{\mathrm{model}}\left(z_1, z_2, \dots, z_N; n_1, n_2, \dots, n_N\right) \propto \\
	\psi_{\mathrm{model}}\left(\frac{c}{z_1}, \frac{c}{z_2}, \dots, \frac{c}{z_N}; n_1, n_2, \dots, n_N\right).
	\label{eq:demand1}
\end{multline}
We demand proportionality instead of equality because the wavefunctions are unnormalized. 

Under the transformation \eqref{eq:inversion}, we have
\begin{equation}
	z_j-z_k\rightarrow \frac{c}{z_j}-\frac{c}{z_k}=-c\,\frac{z_j-z_k}{z_jz_k},
\end{equation}
and 
\begin{multline}
	\prod_{j<k}(z_j-z_k)^{\gamma_j\gamma_k} \rightarrow \\
	\rightarrow \prod_{j<k}(-c)^{\gamma_j\gamma_k} \prod_{j<k}\left(z_jz_k\right)^{-\gamma_j\gamma_k}\prod_{j<k}\left(z_j-z_k\right)^{\gamma_j\gamma_k}.
\end{multline}
Let us look closer at the arising factors. We have
\begin{multline}\label{ctransform}
	\prod_{j<k}(-c)^{\gamma_j\gamma_k}=\prod_{j\neq k}(-c)^{\gamma_j\gamma_k/2}=\\=
	\prod_{j,k}(-c)^{\gamma_j\gamma_k/2}\prod_{j}(-c)^{-\gamma_j^2/2}=\\=
	(-c)^{\frac{\left(p_++p_-\right)^2-\left(q^2-2q\eta\right)M-\eta^2N}{2q}},
\end{multline}
where we used the facts that $\sum_j\gamma_j=-(p_++p_-)/\sqrt{q}$ (from the charge neutrality condition) and that $n_j^2=n_j$ as $n_j\in\{0,1\}$. Because $M$ and $N$ are constant, this term can be absorbed into normalization, so we can ignore it. Similarly,
\begin{multline}
	\prod_{j<k}(z_jz_k)^{-\gamma_j\gamma_k}=
	\prod_{j,k}(z_jz_k)^{-\gamma_j\gamma_k/2}\prod_{j}z_j^{\gamma_j^2}=\\=
	\prod_{j,k}z_j^{-\gamma_j\gamma_k/2}\prod_{j,k}z_k^{-\gamma_j\gamma_k/2}\prod_{j}z_j^{qn_j^2-2\eta n_j+\eta^2/q
	}=\\=
	\prod_{j}z_j^{\gamma_j \frac{p_++p_-}{2\sqrt{q}}}\prod_{k}z_k^{\gamma_k\frac{p_++p_-}{2\sqrt{q}}}\prod_{j}z_j^{(q-2\eta)n_j+\eta^2/q}=\\
	=\prod_{j}z_j^{(q-2\eta+p_++p_-) n_j  -\eta (p_++p_--\eta)/q}.
	\label{eq:zz}
\end{multline}
Thus, the transformed wavefunction coefficients are proportional to
\begin{multline}
	\psi_{\mathrm{model}}\left(\frac{c}{z_1}, \frac{c}{z_2}, \dots, \frac{c}{z_N}; n_1, n_2, \dots, n_N\right)\propto\\
	\delta(qM+p_{+}+p_{-}-N\eta)\prod_{j}\chi_j^{n_j}   
	\prod_{j}z_j^{(q-2\eta+p_+) n_j} \times \\ 
	\prod_{j<k}(z_j-z_k)^{\gamma_j\gamma_k}.
	\label{eq:laughlin_cylinder_transformed}
\end{multline}
By comparing the wavefunctions before and after the transformation, we find that \eqref{eq:demand1} is true if the condition \eqref{eq:ppluspminus} is enforced.

Equation \eqref{eq:demand1} says that the wavefunction is the same if we put the $j$th site at $z_j$ for all $j$ or we put the $j$th site at $c/z_j$ for all $j$. Here, we are, of course, interested in the case, where the lattice itself has inversion symmetry such that the inversion maps the set of lattice coordinates to itself. We choose the numbering of the sites such that the inversion maps site $j$ into site $N+1-j$, i.e.,
\begin{equation}
    z_j=\frac{c}{z_{N+1-j}},
\end{equation}
to avoid complications with branch cuts. We have not yet shown that the wavefunction is invariant under inversion, as the wavefunction does not map to itself under inversion, but rather to the wavefunction in which the sites are numbered from $N$ to $1$ rather than from $1$ to $N$. 

The only part of the wavefunction that is affected by the ordering of the coordinates is the factor $\prod_{j<k}(z_j-z_k)^{\gamma_j\gamma_k}$, which transforms as
\begin{multline}
    \prod_{j<k}(z_j-z_k)^{\gamma_j\gamma_k} \to\\
    \to\prod_{j>k}(z_j-z_k)^{\gamma_j\gamma_k}
    =\prod_{j<k}(-1)^{\gamma_j\gamma_k}\prod_{j<k}(z_j-z_k)^{\gamma_j\gamma_k}
\end{multline}
when we reverse the ordering of the numbering. It follows from \eqref{ctransform} with $-c$ replaced by $e^{i\pi}$ that $\prod_{j<k}(-1)^{\gamma_j\gamma_k}$ is a global phase factor, and hence reversing the ordering of the numbering leaves the wavefunction unchanged up to a global phase factor. Combining this result with \eqref{eq:demand1}, we conclude that if we number the sites such that inversion maps $z_j$ to $z_{N+1-j}$, then the wavefunction is invariant under inversion up to a global factor that can be absorbed in the normalization constant.
	
This reasoning can be generalized to the case of a wavefunction with anyons, if the anyons are distributed in an inversion-symmetric way (that is, there are pairs of anyons with the same $p_k$s, located at $w_k$s being related by inversion symmetry).
	
Finally, we comment on different choices of $\eta$, $p_+$ and $p_-$. When \eqref{eq:ppluspminus} is enforced, the charge neutrality relation $qM +p_++p_- - N\eta=0$ of the wavefunction without anyons becomes
\begin{equation}
	q(M-1)+2p_{-}=N\eta-2\eta.
	\label{eq:condition}
\end{equation}
For example, if $p_{-}=0$, then, as $\eta\rightarrow$ 0 at constant $N\eta$, we obtain the relation
\begin{equation}
	q(M-1)=N\eta.
	\label{eq:condition2}
\end{equation}
This is similar to the relation between particle number and flux in the continuum Laughlin state on the sphere, where ``$-1$'' is related to the ``shift'' quantum number \cite{wen1992shift}. 
	
To compare the model lattice Laughlin state to the exact diagonalization results, we set $\eta=\alpha$. Then, \eqref{eq:condition2} is typically not fulfilled, i.e.\ $p_-\neq 0$. If we have a given $\eta$, $M$ and $N$, the correct $p_-$ can be determined from \eqref{eq:condition}, and is equal to \eqref{eq:condition3}.
	
We note that in the above calculations the total flux $N\eta$ is different from the total flux used in Eq.~\eqref{eq:filling_plaq}, i.e.\ $N\eta\neq \alpha N_{\mathrm{plaq}}$. This is because in this Appendix we assign a flux $\eta$ to a given site, while in Eq.~\eqref{eq:filling_plaq} the flux $\alpha$ is assigned to a given plaquette (and we have to set $\eta=\alpha$ so that the flux encircled when going around a plaquette is the same in both cases).

\begin{figure}
	\includegraphics[width=0.5\textwidth]{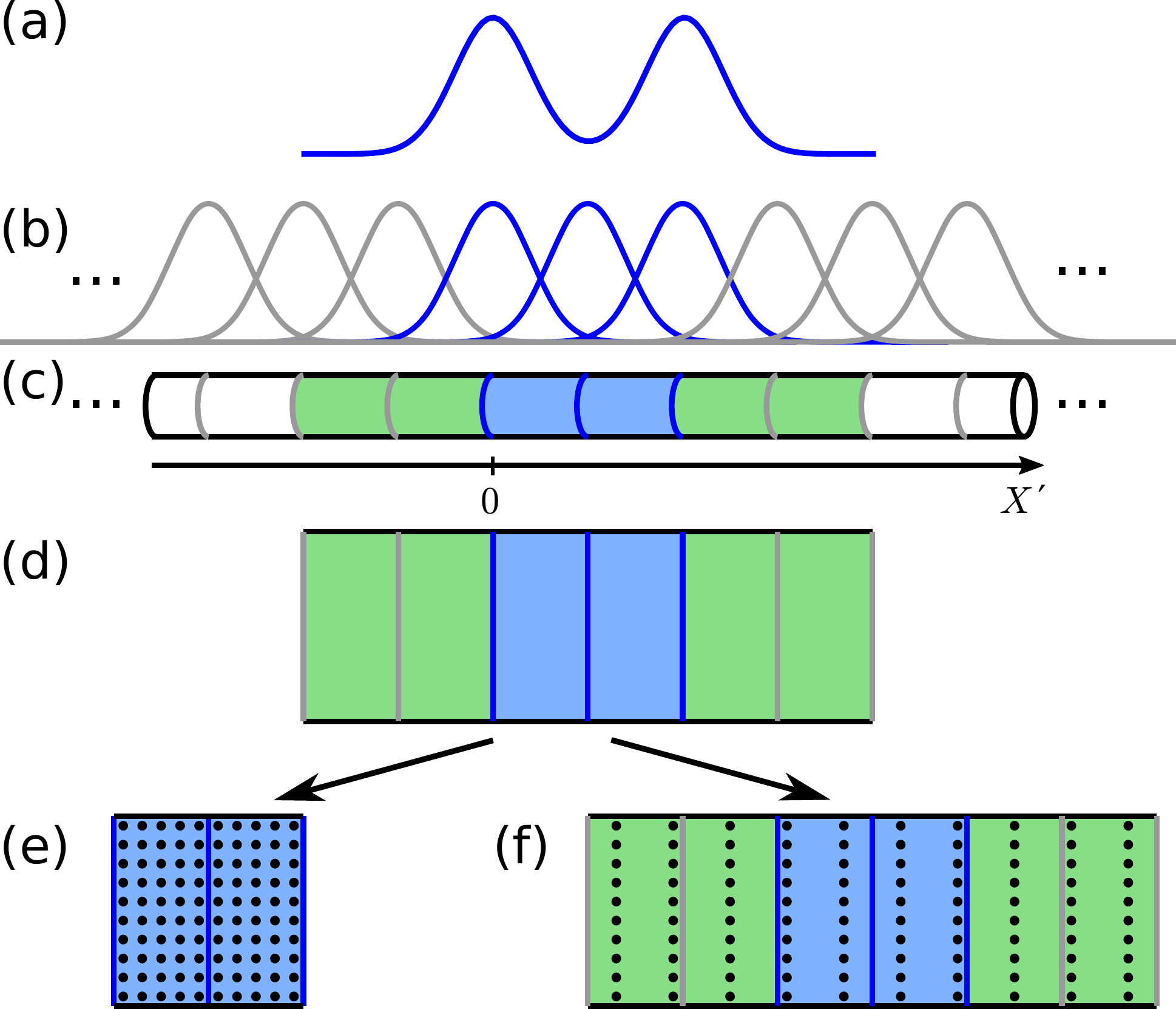}
	\caption{The relation between continuum and lattice systems for the example with $q=2$, $M=2$, and $L_y'=2\sqrt{\pi}l_B'$ described in the main text. The many-body wavefunction, with $x$-dependent particle density shown schematically in (a), is composed out of three single-particle orbitals with $m=0,1,2$ -- the blue ones in (b). The cylinder is drawn schematically in (c), with blue and grey lines denoting the Gaussian centers of the orbitals. The blue region of the cylinder located between the centers of the 0th and 2nd orbitals contains $q(M-1)=2$ flux quanta. The dots ``$\dots$'' in (b) and (c) mean that the cylinder extends infinitely in both positive and negative $x$ directions. The blue and green regions contain a total number of 6 flux quanta. In (d) we show the blue and green regions unwound to a plane. The circumference of the cylinder is chosen in such a way that the blue region is a square. In (e) and (f) we show the site positions for the analogous lattice systems with total flux $N\eta=2$ and $N\eta=6$, respectively, for an $N_x\times N_y=10\times 10$ lattice.}
	\label{fig:CylinderSchematic}
\end{figure}
	
\begin{figure}
	\includegraphics[width=0.5\textwidth]{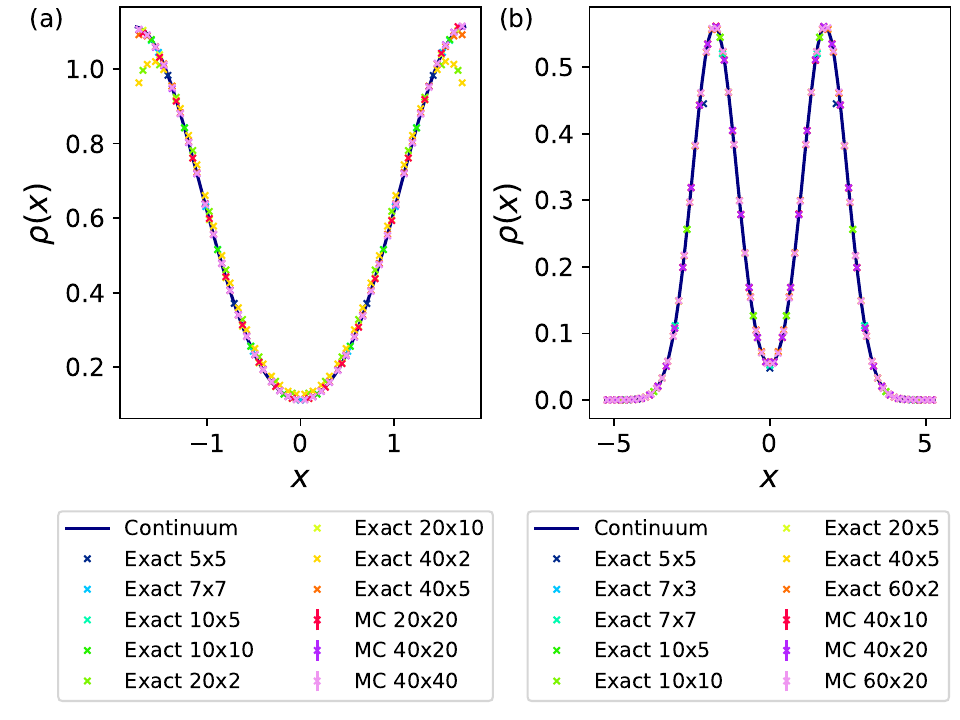}
	\caption{Comparison of the rescaled particle density for the example with $q=2$, $M=2$, and $L_y'=2\sqrt{\pi}l_B'$ described in the main text for the total flux of (a) two flux quanta and (b) six flux quanta. The $x$ coordinate is measured in units of the magnetic length of the given system.}
	\label{fig:LatticeVsContinuum}
\end{figure}

\section{Relation between lattice and continuum particle densities}\label{app:LatticeVsContinuum}
	
In the following, we will compare the particle density in the lattice wavefunction \eqref{eq:laughlin_cylinder} to the case of the continuum Laughlin wavefunction. This can serve as a sanity check for \eqref{eq:laughlin_cylinder}, as in the limit $\eta\rightarrow 0$ we expect it to converge to the continuum wavefunction. Also, it connects \eqref{eq:laughlin_cylinder} to the commonly used concepts such as single-particle Landau level orbitals.
	
The continuum Laughlin wavefunction \cite{rezayi1994laughlin} is given by
\begin{multline}
	\Psi(\Xi_1', \Xi_2', \dots \Xi_M')=\\=
	\prod_{j<k}(Z_j'-Z_k')^{q}\exp\left(-\frac{\sum_jX_j'^2}{2l_B'^2} \right),
	\label{eq:cont_wfn}
\end{multline}
where $\Xi_j'=X_j'+iY_j'$ are the coordinates of the particles, $Z_j'=\exp(2\pi \Xi_j'/L_y')$, and $l_B'$ is the magnetic length. We use primed symbols for the lengths and coordinates in the continuum system to stress that in principle they can be measured in different units than in the lattice case. A schematic plot of particle density as a function of $X'$  in the case of $q=2$, $M=2$ and $L_y'=2\sqrt{\pi}l_B'$ is shown in Fig.\ \ref{fig:CylinderSchematic}(a).
	
It is instructive to view \eqref{eq:cont_wfn} in terms of single-particle lowest Landau level orbitals on a cylinder \cite{rezayi1994laughlin}. These orbitals are labeled by an integer $m$ and given by
\begin{multline}
	\phi_m(X',Y')=\exp\left(2\pi i \frac{m Y'}{L_y'}\right) \times \\ \times
	\exp\left(-\frac{1}{2l_B'^2}\left(X'- \frac{2\pi ml_B'^2}{L_y'}\right)^2\right),
\end{multline}
which consist of a plane wave in the $y$ direction and a Gaussian centered at $\frac{2\pi ml_B'^2}{L_y'}$ in the $x$ direction. The density profiles of these orbitals, as a function of the $X'$ coordinate, are shown schematically in Fig.\ \ref{fig:CylinderSchematic}(b), and the position of their Gaussian centers on the cylinder (the grey and blue rings) in Fig.\ \ref{fig:CylinderSchematic}(c). The basis for \eqref{eq:cont_wfn} is constructed out of orbitals with $m=0, 1, \dots, q(M-1)$.
	
The region between the Gaussian centers of the nearest-neighbor orbitals always contains one flux quantum. Therefore, the $x$ distance between the Gaussian centers depends on $L_y'$. As the circumference of the cylinder decreases, the orbitals get more separated, and the nearly-uniform quantum Hall state transforms into a charge density wave \cite{rezayi1994laughlin} (in the example in Fig.\ \ref{fig:CylinderSchematic}(a), the inhomogenity of the density is already quite developed, with high density on the $m=0,2$ orbitals and low density on the $m=1$ orbital). 
	
We also note that \eqref{eq:cont_wfn} is in fact defined on an infinite cylinder. However, we can also restrict \eqref{eq:cont_wfn} to a finite cylinder by truncating the domain of $\Xi_j'$ to some finite range and normalizing the wavefunction appropriately. One possible choice is to set the cylinder borders to the centers of $m=0$ and $m=q(M-1)$ orbitals -- in such a way the system is pierced by $q(M-1)$ flux quanta, in accordance with \eqref{eq:condition2}. In our $M=2$ example, this is the blue region in Fig.~\ref{fig:CylinderSchematic}(c), also shown in Fig.~\ref{fig:CylinderSchematic}(d) after unwinding to a plane. It contains two flux quanta. However, we can also use a bigger system, which would allow us to observe the Gaussian ``tails'' at the edges. For example, we can study a system thrice as large (six flux quanta), denoted by combined green and blue regions in Fig.\ \ref{fig:CylinderSchematic}(c), (d). Since we focus on the systems where the particle density is inversion-invariant, we consider only cases where the lengths added before the center of $m=0$ orbital and after the center of $m=q(M-1)$ orbitals are the same (i.e.\ that the two green regions in Fig.~\ref{fig:CylinderSchematic}(c), (d) have the same size). We do not restrict the total flux to be an integer, i.e.\ the edges of the domain do not have to lie at the Gaussian centers as in Fig.\ \ref{fig:CylinderSchematic}.
	
Now, knowing the meaning of various parameters of the continuum wavefunction, we propose that to match the particle densities of continuum and lattice wavefunctions, we have to do the following:
\begin{enumerate}
	\item obtain the magnetic length $l_B$ of the lattice system to establish a common length scale,
	\item match the circumferences $L_y'/l_B'=L_y/l_B$ (i.e.\ the distance between the Gaussian centers),
	\item match the total flux through the system (i.e.\ the extent of the system in the $x$ direction),
	\item match the center of symmetry (i.e.\ shift the $x$ coordinates so that the center of both systems is at $x=0$),
\end{enumerate}
not necessarily in this order (it depends on which parameters of which system we want to vary).
	
As an example, let us consider finding lattice systems corresponding to the $M=2$ continuum wavefunction from Fig.\ \ref{fig:CylinderSchematic} defined on the blue region (two flux quanta). We will consider a rectangular lattice of a fixed size $N_x \times N_y$ in terms of unit cells, with a unit cell of size $a \times 1$, where $a$ will be adjusted in the process. We have $N=N_xN_y$ and $L_y=N_y$. Thus, by doing step 2 (i.e.\ demanding $L_y'/l_B'=L_y/l_B$), and recalling that for the system from Fig.\ \ref{fig:CylinderSchematic} we had $L_y'=2\sqrt{\pi}l_B'$, we find $l_B=N_y/(2\sqrt\pi)$. From step 3 we get that $N\eta=2$, and hence $\eta=2/N$. Now, let us look again at the magnetic length. Eq.~\eqref{eq:laughlin_cylinder} does not refer to it explicitly, but we can infer the relation of $l_B$ to the unit cell size by considering a loop and comparing its area to the encircled flux (for simplicity, we treat the flux as uniformly distributed in space). A unit cell of size $a\times 1$ corresponds to $\eta$ flux quanta, so we must have $a=2\pi \eta l_B^2$. In this way, we determined the shape of the lattice.
	
Next, we find $p_-$ from \eqref{eq:condition3}. Figure \ref{fig:LatticeVsContinuum}(a) shows the comparison of the particle densities for continuum and lattice for various $N_x \times N_y$. In the plot, the $x$ coordinate is plotted in units of magnetic length of the respective system and shifted so that the center of the system is at $x=0$. The particle density is normalized so that its integral is equal to $M$ (in the case of lattice systems, we plot $\rho(x)=\langle n(x) \rangle l_B/a$). The densities are calculated exactly for sufficiently small systems and using Monte Carlo for the largest ones. It can be seen that there is a good match between the lattice and continuum wavefunctions, especially in the center. Notable mismatches, especially near the edges, exist only for the cases with small $N_y$.  As $N_y$ approaches $N_x$, the mismatch vanishes almost completely. 
	
Similarly, we can repeat the procedure for the combined blue and green regions of Fig.\ \ref{fig:CylinderSchematic} (six flux quanta), with the result shown in Fig.\ \ref{fig:LatticeVsContinuum}(b). Again we obtain a good match, with small but visible mismatches for small $N_x$. While in the case of two flux quanta and small $N_y$ the density near the edges significantly departed from the continuum values, here we do not observe such an effect, probably because the system is more elongated, or because the density at the edges is very small.
	
If one considers a Hofstadter model \eqref{eq:BH_1} with the gauge \eqref{eq:landaugauge}, then one can compare the particle densities in the ground state and the model wavefunction in an analogous way. Matching the wavefunctions to compute the overlap is also possible, but more complicated. The model wavefunction has to be centered at the $m=0$ orbital, which can be achieved by a shift of the $X'_j$ coordinates together with a modification of the phase of \eqref{eq:cont_wfn} (the phase transformation does not affect the particle density).
	

%

\end{document}